\documentstyle[prl,aps,epsf,twocolumn]{revtex}
\input epsf
\begin{document}
\title{Pascal Principle for Diffusion-Controlled Trapping Reactions.
}
\author{
\parindent=0.0in
{M.Moreau,$^1$ G.Oshanin,$^1$
O.B{\'e}nichou,$^2$ M.Coppey $^1$}
{\\
\textit{$^1$ Laboratoire de Physique Th{\'e}orique des Liquides, 
University Pierre et Marie Curie, 75252 Paris Cedex 05, France \\
\vspace*{.1in}
$^2$ Laboratoire de Physique de la Mati\`ere Condens\'ee, 
Coll{\`e}ge de France, 11 place Marcelin Berthelot, 75005, Paris, France}\\
}
(Submitted on December 20, 2002)\\
(Accepted by PRE on March 13, 2003)\\
\parbox{14cm}{
In this paper we analyse the long-time behavior of the 
survival probability $P_A(t)$
of an $A$ particle,
which performs lattice random walk in the presence of
randomly moving traps $B$. We 
show that for both perfect and 
imperfect trapping reactions, 
for arbitrary spatial dimension $d$ and for a rather
general class
of random walks, $P_A(t)$ is less than 
or equal 
to 
the survival probability of an $\it immobile$ target $A$ in 
the presence of randomly moving traps. 
{\flushleft PACS numbers: 05.40.-a, 02.50Ey, 82.20.-w \hspace*{\fill}}
}
\normalsize
}
\maketitle

Blaise Pascal has 
once asserted that all 
misfortune of man comes from the fact that 
he does not stay peacefully in his room [1].  
Taking this statement 
out of its philosophical context, 
it might be tempting to 
evoke it 
as the "Pascal principle" 
in regard to 
the problem of 
survival 
of an $A$ 
particle, which moves 
randomly in a sea
of randomly moving 
traps $B$, 
(presented at mean density $"b"$), 
and
is 
annihilated upon the first encounter
with any of them. 
This complex problem, which is
intimately related to many fundamental problems
of statistical physics, 
has been attracting a great deal of 
attention within the last three decades
(see, e.g., Refs.\cite{4,21,23} 
and references therein).
Many important results have been obtained 
but an exact 
solution is lacking as yet.

In a recent paper \cite{2}, which  focused 
on 
the behavior in the particular case
when both $A$ and $B$s perform conventional diffusive motion,
it has been 
claimed that it is 
intuitively clear
that when the traps are 
initially 
(statistically)
symmetrically placed with respect 
to an $A$ particle, 
the latter
will on average survive longer if it 
stays still than if it diffuses. 
This represents,
if true, a clear illustration 
of the 
"Pascal principle", and implies 
that the $A$ particle 
survival probability $P_A(t)$ 
obeys the inequality
\begin{equation}
\label{ineq}
P_A(t) \leq P'_A(t),
\end{equation}
where $P'_A(t)$ stands for 
the corresponding
survival probability 
of an $\it immobile$ target $A$ in 
presence of diffusive
traps. The latter 
can be  
evaluated exactly 
\cite{4,3}. 
Furthermore,
following the reasonings of 
earlier works \cite{4,5}, 
the authors of Ref.\cite{2}
constructed 
a lower bound
on $P_A(t)$ and
showed that
these bounds converge
as $t \to \infty$ for 
systems of spatial 
dimension $d \leq 2$, 
defining thus 
the large-$t$ asymptotic 
form of $P_A(t)$ exactly!
Subsequently, in Ref.\cite{6}, the arguments 
of Ref.\cite{2} have been generalized
to arbitrary symmetric random motion 
with particles' and traps'
trajectories
characterized 
by a fractal dimension 
$d_\omega$ (not necessarily equal to $2$, which is specific for conventional diffusive motion), 
and here  
exact 
large-$t$ asymptotic forms of 
$P_A(t)$ have been obtained
for systems of spatial dimension 
$d \leq d_\omega$. 

We note that the inequality 
in Eq.(\ref{ineq})
has been derived previously 
for the process of 
hopping transport 
of an excitation
on a disordered array of 
immobile
donor centers in presence of 
randomly placed, immobile 
quenchers \cite{burl}. 
On the other hand, Eq.(\ref{ineq})
is compatible 
with recent 
results on
 ballistic 
$A + A \to 0$ annihilation
process \cite{jaroslaw}.
For trapping $A +  B \to B$ reactions (TR)
involving $\it diffusive$ species 
the authors of Ref.\cite{2} 
were unable, 
however, to
prove 
the inequality in Eq.(\ref{ineq}), 
but  furnished only some 
arguments in favor of it.
Consequently, results on $\it exact$ 
asymptotic behavior of $P_A(t)$ depend crucially 
on whether
the inequality in 
Eq.(\ref{ineq}) is indeed correct.

In this paper 
we analyse, in the lattice formulation of the model,
the 
Pascal principle-like inequality 
in Eq.(\ref{ineq}). 
Following the line of argument 
of Refs.\cite{3} and \cite{mont},
we show that
for both instantaneous (perfect) and imperfect 
TRs, 
for arbitrary spatial dimension $d$
and for a rather general class of random walks 
(not necessarily conventional
diffusion), 
$P_A(t)$ of 
a mobile
$A$ particle 
in presence of mobile traps, 
is less than the  
survival probability of an 
immobile target $A$ in 
the presence of mobile 
$B$s.  Our proof assumes essentially that $B$s
perform a space- and time-homogeneous, unbiased jump process on the lattice sites. 
Behavior in the continuous-space systems, 
which requires a somewhat more delicate
analysis (especially for $d \geq 2$), 
will
be presented
elsewhere \cite{7}.

Consider a $d$-dimensional hypercubic lattice 
containing $M$ sites. 
A single $A$ particle is initially 
located 
at the origin, while $N$ traps $B$ are 
placed on the lattice 
at 
positions ${\bf Y}_0^{(i)} \neq 0$, where the superscript $i$
here and henceforth
numerates the traps, $i = 1, \ldots ,N$.

In regard to particles dynamics,
we suppose 
that the $A$ particle
performs a continuous-time 
jump process on the lattice sites
and that the time interval 
between the consecutive
jumps is a 
random variable.
We denote then as $\Gamma_A$ the $A$ particle trajectory 
recorded at integer 
time moments $k = 0, \ldots ,n$,
such that $\Gamma_A = \{{\bf X}_0 = 0, {\bf X}_1, \ldots , {\bf X}_n\}$,
where ${\bf X}_k$ is the vector of a lattice site 
at which the $A$ particle 
resides 
at time moment $k$. Note that
since the time interval between the 
consecutive 
jumps is a random variable, two successive positions 
are not necessarily
different, and not necessarily nearest neighbors.

Now, we suppose that the $B$s 
perform identical and 
independent \textit{discrete time 
random walks}: that is, at each tick of the clock
each $B$ 
can jump with a given probability 
from a 
lattice site ${\bf Y}$ to another (not necessarily neighboring) 
site ${\bf Y'}$, or it may also remain 
at ${\bf Y}$. 
We define then as $\Gamma_B^{(i)}$ the 
trajectory of the $i$-th $B$ particle,
$\Gamma_B^{(i)} = \{{\bf Y}_0^{(i)}, {\bf Y}_1^{(i)}, \ldots , {\bf Y}_n^{(i)}\}$,
where ${\bf Y}_k^{(i)}$ 
denotes 
the position of the $i$-th trap 
at time moment $k$, 
$k = 0, 1, \ldots, n$. 

Next, let $P({\bf Y}_n^{(i)}|{\bf Y}_0^{(i)})$
be the conditional probability of 
finding the $i$-th trap
B at site ${\bf Y}_n^{(i)}$ 
at time moment $n$, knowing that it started its random walk 
at ${\bf Y}_0^{(i)}$.
We assume now 
that random walks executed by the $B$ particles 
satisfy the following, quite general conditions:\\
({\bf i})  the random walk is space- and time-homogeneous:  
\begin{equation}
P({\bf Y}_n^{(i)}|{\bf Y}_0^{(i)}) = 
P({\bf Y}_n^{(i)} - {\bf Y}_0^{(i)}|0),
\end{equation}
({\bf ii}) at any time moment 
$n$ and for any $i$ \cite{remark}: 
\begin{eqnarray}
\label{jj}
\displaystyle
P({\bf Y}_n^{(i)} \neq {\bf Y}_0^{(i)}|{\bf Y}_0^{(i)}) \leq 
 P({\bf Y}_n^{(i)} = {\bf Y}_0^{(i)}|{\bf Y}_0^{(i)})  
\equiv R_n,	
\end{eqnarray}
i.e., the conditional probability 
$P({\bf Y}_n^{(i)} \neq {\bf Y}_0^{(i)}|{\bf Y}_0^{(i)})$ of 
finding at time moment 
$n$ the $i$-th trap
at site $Y_n^{(i)}$ different 
from its starting point $Y_0^{(i)}$,  
is less than or equal to 
the probability $P({\bf Y}_n^{(i)} = {\bf Y}_0^{(i)}|{\bf Y}_0^{(i)})$
of finding it at time moment $n$ exactly 
at 
the starting point;
here,
$R_n$ denotes the return probability 
of the random walk executed by the traps. By convention, $R_0 \equiv 1$.

Finally, we 
consider two situations with respect to reaction; namely,
when a) the $A$ particle 
gets annihilated  
with probability $p = 1$ upon the first encounter
with any of $B$s (perfect reaction) and b)
when the annihilation of the $A$ particle
takes place with probability $p < 1$
when an $A$ and any of $B$s occur at the same site (imperfect reaction). 
For computational convenience,
we stipulate that for both situations 
reaction can take place only at integer time moments; that is,
if 
at a non integer time $A$ 
jumps on a site 
which is occupied by any $B$, $A$ 
survives till the 
departure of this particle or 
an arrival of another $B$. 
The probability that both $A$ and $B$ particles jump on 
the same site simultaneously is clearly equal to zero.

Let 
$P^{(i)}_n(\Gamma_A|{\bf Y}_0^{(i)})$ denote 
\textit{the 
conditional probability} that 
for a given realization of the $A$ particle 
trajectory $\Gamma_A$, 
the $i$-th $B$ particle starting its walk from the point ${\bf Y}_0^{(i)}$
\textit{does not destroy} (encounter)  
$A$ \textit{up to time} $n$.
Since $B$s move  and act
independently of each other, 
the conditional 
probability $\Psi_n(\Gamma_A|\{{\bf Y}_0^{(i)}\})$ that, for a
 given realization of the $A$ particle
 trajectory $\Gamma_A$
and a given set of the starting points 
$\{{\bf Y}_0^{(i)}\}$,
the
$A$ particle survives up to time $n$, is 
determined by
\begin{eqnarray}
\label{1}
\displaystyle
\Psi_n(\Gamma_A|\{{\bf Y}_0^{(i)}\}) & = 
& \prod_{i=1}^N P^{(i)}_n(\Gamma_A|{\bf Y}_0^{(i)}),
\end{eqnarray}
and hence, the $A$ particle 
survival probability obeys
\begin{eqnarray}
\label{2}
\displaystyle
P_A(n) & = & \left\langle \Big\langle \Psi_n(\Gamma_A|\{{\bf Y}_0^{(i)}\})
 \Big\rangle_{\{{\bf Y}_0^{(i)}\}}\right\rangle_{\Gamma_A},
\end{eqnarray}
the average being taken first over the starting points of $B$s and then over
all possible trajectories $\Gamma_A$.  

Now, supposing that $B$s were 
initially uniformly distributed
on the lattice (excluding the origin) and droping the superscript $"i"$,
one finds
from Eqs.(\ref{1}) and (\ref{2}) that
\begin{eqnarray}
\displaystyle
P_A(n) & = & \left\langle \left\{ \frac{1}{M} \sum_{{\bf Y}_0\neq0} 
P_n(\Gamma_A|{\bf Y}_0) \right\}^N  \right\rangle_{\Gamma_A} = \nonumber\\
& = &\left \langle \left\{ 1 - \frac{1}{M} \sum_{{\bf Y}_0 \neq 0} 
\Big( 1-P_n(\Gamma_A|{\bf Y}_0) \Big) \right \}^N \right \rangle_{\Gamma_A}.
\end{eqnarray}
Turning next to the thermodynamic limit, i.e. setting
$N,M = \infty$ with a fixed ratio $b = N/M$,
we find 
\begin{eqnarray}
\label{4}
\displaystyle
P_A(n) & = & \left\langle \exp{\left\{-b\sum_{{\bf Y}_0\neq0}
 \Big( 1-P_n(\Gamma_A|{\bf Y}_0) \Big)\right\}} \right\rangle_{\Gamma_A}.
\end{eqnarray}
Consequently, the survival probability $P_A(n)$ 
can be thought of as the generating function
of the probability
\begin{eqnarray}
\label{fg}
\displaystyle
Q_n(\Gamma_A|{\bf Y}_0) & = &  1 -  P_n(\Gamma_A|{\bf Y}_0)
\end{eqnarray}
that for a given $\Gamma_A$,
a single $B$, being at ${\bf Y}_0 \neq 0$ at $n = 0$,
destroys the $A$ at some 
time moment $\leq n$. 
Note also that when the $A$ particle
is immobile, Eq.(\ref{4}) reduces 
to
\begin{eqnarray}
\label{40}
\displaystyle
P'_A(n) & = & \exp{\left\{-b\sum_{{\bf Y}_0\neq0} \Big( 1-P_n(0|{\bf Y}_0) \Big)\right\}},
\end{eqnarray}
which can be evaluated
explicitly
 \cite{4,3}. 

We seek now an upper bound on the survival probability in Eq.(\ref{4}).
Let $F_k(\Gamma_A|{\bf Y}_0)$ denote 
\textit{the conditional probability that} a single $B$ particle, 
being at ${\bf Y}_0$ at $k=0$, 
encounters $A$  
for the first time at time moment 
$k$, 
given the $A$ particle  trajectory $\Gamma_A$ is fixed.
Then, the conditional probability 
$Q_n(\Gamma_A|{\bf Y}_0)$  
that a single $B$ particle, 
starting from 
${\bf Y}_0$, destroys $A$ at or before $k = n$ obeys
\begin{eqnarray}
\label{gf}
\displaystyle
Q_n(\Gamma_A|{\bf Y}_0) & = &  \sum_{0\leq k\leq n} F_k(\Gamma_A|{\bf Y}_0).
\end{eqnarray}
Now, the conditional probability that the trajectory of 
$B$ (extended after the possible annihilation 
of $A$) meets $\Gamma_A$ at time $n$ (not necessarily for 
the first time) is clearly
\begin{eqnarray}
\displaystyle
\nonumber P({\bf Y}_n &=& {\bf X}_n| {\bf Y}_0) = F_n(\Gamma_A|{\bf Y}_0) +  \\
&+& \sum_{0\leq k<n} P({\bf Y}_n = {\bf X}_n| {\bf Y}_k) F_k(\Gamma_A|{\bf Y}_0). 
\end{eqnarray}
Summing both sides of 
the last 
equation over 
all initial positions ${\bf Y}_0\neq 0$ we obtain
\begin{eqnarray}
\label{t}
\displaystyle
\nonumber \sum_{{\bf Y}_0\neq0}P({\bf Y}_n &=& {\bf X}_n| {\bf Y_0}) = {\it K}_n(\Gamma_A) + \\
&+& \sum_{0\leq k<n} P({\bf Y}_n = {\bf X}_n| {\bf Y}_k) {\it K}_k(\Gamma_A),
\end{eqnarray}
where ${\it K}_n(\Gamma_A)$ is defined as
\begin{eqnarray}
\label{zu}
\displaystyle
{\it K}_n(\Gamma_A) & = & \sum_{{\bf Y}_0\neq0} F_n(\Gamma_A|{\bf Y}_0).
\end{eqnarray}
We note that ${\it K}_n(\Gamma_A)$ has a meaning of a time-dependent reaction rate;
using Eq.(\ref{zu}) we rewrite 
Eq.(\ref{4}) as
\begin{eqnarray}
\label{44}
\displaystyle
P_A(n) & = & \left\langle \exp{\left\{ - b \sum_{0\leq k\leq n}  {\it K}_k(\Gamma_A)
\right\}} \right\rangle_{\Gamma_A}.
\end{eqnarray}
On the other hand, the survival probability $P'_A(n)$
of an $\it immobile$ $A$ particle, Eq.(\ref{40}), 
can be written as
\begin{equation}
\label{as}
P'_A(n) = \exp{\left\{- b \sum_{0 \leq k \leq n} {\it K}_k\right\}} = \exp{\Big\{- b \left(S_n - 1\right)\Big\}},
\end{equation}
where
${\it K}_k \equiv {\it K}_k(\Gamma_A \equiv 0)$, while $S_n$
is the expected number of distinct 
sites visited by a single $B$ up to time moment $n$ (see, 
e.g., Ref\cite{nous} for more details). 
The last quantity is obtained directly 
by inversion of
its generating function $\widehat{S} = \sum_{n=0}^{\infty} S_n \xi^n$,
which can be evaluated 
explicitly,
$\widehat{S} = (1-\xi)^{-2} \widehat{R}^{-1}$, $\widehat{R} = \sum_{n=0}^{\infty} R_n \xi^n$
being the generating function of the return probability $R_n$.

We turn now to the comparison of ${\it K}_k(\Gamma_A)$ and ${\it K}_k$.
Using the normalization 
$\sum_{{\bf Y}_0}P({\bf Y}_n = {\bf X}_n| {\bf Y}_0)=1$,
and the condition (i), 
we have
$\sum_{{\bf Y}_0 \neq 0} P({\bf Y}_n = {\bf X}_n| {\bf Y}_0) = 1-P({\bf Y}_n 
= {\bf X}_n| 0)$.
Consequently, in virtue of (ii):
\begin{equation}
\label{j2}
\sum_{{\bf Y}_0\neq 0} P({\bf Y}_n = {\bf X}_n|{\bf Y}_0) \geq 1 - R_n.
\end{equation}
Now, from Eqs.(\ref{t}) and
(\ref{j2}) we get
\begin{equation}
\label{ff}
{\it K}_n(\Gamma_A) + \sum_{0 \leq k < n} P({\bf Y}_n = {\bf X}_n|{\bf Y}_k) {\it K}_k(\Gamma_A) \geq 1 - R_n.
\end{equation}
On the other hand, the inequality in Eq.(3) implies that
\begin{equation}
\label{gg}
P({\bf Y}_n = {\bf X}_n|{\bf Y}_k) \leq R_{n-k}.
\end{equation}
Recollecting that $R_0 = 1$ and making use of Eq.(\ref{gg}), we thus
enhance the inequality in Eq.(\ref{ff}), 
which now reads
\begin{eqnarray}
\label{nnn}
\displaystyle
\sum_{0\leq k \leq n} R_{n-k}{\it K}_k(\Gamma_A) \geq 1 - R_n,
\end{eqnarray}
and also becomes an equality when 
$A$ is $\it immobile$.

Further on, multiplying both sides of Eq.(\ref{nnn}) by $\xi^n$
and performing summation over $n$, we get
\begin{equation}
\label{it}
\widehat{\it K}(\Gamma_A) = \sum_{n = 0}^{\infty} {\it K}_n(\Gamma_A) \xi^n \geq \frac{1}{(1-\xi) \widehat{R}} - 1.
\end{equation}
Next, taking into account that
\begin{equation}
\widehat{S}(\Gamma_A) = \sum_{n=0}^{\infty} \xi^n 
\left(\sum_{0 \leq k \leq n} {\it K}_k(\Gamma_A)\right) = \frac{\widehat{\it K}(\Gamma_A)}{1-\xi},
\end{equation}
we find from Eq.(\ref{it}) the following inequality: 
\begin{equation}
\widehat{S}(\Gamma_A) \geq \frac{1}{(1-\xi)^2 \widehat{R}} - \frac{1}{1-\xi} 
\equiv \sum_{n =0}^{\infty} \xi^n \left(\sum_{0\leq k \leq n} {\it K}_k\right),
\end{equation}
which implies 
that the generating function of the
expression in the exponent in Eq.(14), describing 
the $A$ particle survival probability
 in case when it "leaves the room" 
and
changes its position with time,
 is always greater 
than or equal to
the generating function
of the expression
in the exponent in Eq.(\ref{as}), 
which applies to the
case when the $A$ particle stays 
peacefully at its initial position.

Hence, turning to the limit $\xi \to 1^{-}$ ($n \to \infty$) 
and making use
of the Tauberian theorems \cite{9}, we arrive at the conclusion that 
the desired inequality in Eq.(1) holds 
in the limit $n \to \infty$. As a matter of fact, it can be shown that 
this inequality holds generally for arbitrary finite $n$; the proof 
in this statement is, however, rather cumbersome and will be presented 
elsewhere 
\cite{7}.

Lastly, we briefly outline the steps
involved in the derivation of Eq.(1) in the general case 
when reaction between an $A$ and any of $B$s
is not instantaneous,
but takes place with some finite probability $p$.
Following Ref.\cite{nous}, 
we 
suppose
that here 
each trap bears "a gate", which may be 
either open or closed; in the former 
case the trap 
is reactive and
annihilates the $A$ particle
upon the encounter, while
in the latter case
it is inert with respect 
to reaction.
The state of the gate on the $i$-th trap 
is
characterized by a 
random variable $\zeta_i$ such that 
$\zeta_i = 1$ (open gate) 
with probability $p$, and $\zeta_i = 0$ (closed gate)
with the probability $1 - p$, respectively. 
Each $\zeta_i$ updates 
its state at each tick of the clock;
the updating process proceeds completely at random, 
without memory in time and without
correlations with the gates 
imposed on other $B$ particles.
As shown in Ref.\cite{nous}, such a model with stochastic, two-state gates
corresponds to
situations in which 
the elementary reaction act 
is characterized by a finite intrinsic 
reaction constant $K_{el} = p/(1-p)$.

Now,  
we notice that $P_A(n)$
in this case
can be still 
written in the form of Eq.(\ref{4})
with $Q_n(\Gamma_A|{\bf Y}_0)$ defined by Eqs.(\ref{fg}) and (\ref{gf}),
but here 
$F_k(\Gamma_A|{\bf Y}_0) = F_k^{(p)}(\Gamma_A|{\bf Y}_0)$ should be interpreted
as 
the conditional 
probability that the $B$ particle 
encounters 
the $A$ particle 
for the first time at time moment 
$k$ exactly and moreover, that at this moment of 
time the $B$ particle is in reactive state; the superscript $"(p)"$
will signify that here we deal with imperfect TR. 
Further on,
let
$P^{(p)}({\bf Y}_n={\bf X}_n|{\bf Y}_0)$
be 
the conditional probability that the 
trajectory of $B$ meets 
$\Gamma_A$ at time $n$ 
(not necessarily for the first time) 
and at this time moment $B$ is in 
reactive state. For the model under study, such a probability 
obeys
\begin{eqnarray}
\displaystyle
P^{(p)}({\bf Y}_n = {\bf X}_n| {\bf Y}_0) & = & p \; P({\bf Y}_n = {\bf X}_n| {\bf Y}_0)
\end{eqnarray}
and
\begin{eqnarray}
\label{ttt}
\displaystyle
P^{(p)}({\bf Y}_n &=& {\bf X}_n| {\bf Y}_0) = F_n(\Gamma_A|{\bf Y}_0) + \nonumber\\
&+&\sum_{0\leq k<n} P^{(p)}({\bf Y}_n = {\bf X}_n| {\bf Y}_k) F_k(\Gamma_A|{\bf Y}_0). 
\end{eqnarray}
Summing both sides 
of Eq.(\ref{ttt}) over ${\bf Y}_0 \neq 0$, we get
\begin{eqnarray}
\displaystyle
\sum_{{\bf Y}_0\neq 0}P^{(p)}({\bf Y}_n = {\bf X}_n| {\bf Y}_0)  = 0)  \geq  p \left(1-R_n\right),
\end{eqnarray}
and hence, we find
\begin{eqnarray}
\displaystyle
p \left(1-R_n\right) & \leq & {\it K}_n^{(p)}(\Gamma_A) + p \sum_{0\leq k<n}R_{n-k}{\it K}_k^{(p)}(\Gamma_A),
\end{eqnarray}
where 
${\it K}_n^{(p)} = \sum_{{\bf Y}_0 \neq 0} F_n^{(p)}(\Gamma_A|{\bf Y}_0)$. 
Multiplying both sides of Eq.(26)  
by $\xi^n$ and then summing it over
$n$, $n =0, \ldots, \infty$, we arrive at the following 
inequality:
\begin{equation}
\widehat{S}^{(p)}(\Gamma_0) \geq \frac{p \left[1 - (1 -\xi)
\widehat{R}\right]}{(1 - \xi)^2 \left(1 - p + p \widehat{R}\right)},
\end{equation} 
where $\widehat{S}^{(p)}(\Gamma_A)$ is 
the generating function of the sum $\sum_{0 \leq k \leq n} {\it K}^{(p)}(\Gamma_A) = b^{-1} \ln\left(1/P^{(p)}_A(n)\right)$, $P_A^{(p)}(n)$
being the $A$ particle survival probability for imperfect TR.

Turning to the limit $\xi \to 1^-$ ($n \to \infty$), we notice 
that here $(1-\xi) \widehat{R} \ll 1$ and hence, in this limit 
only the first term in the square brackets matters. 
On the other hand,
this leading term  $ \widehat{S}^{(p)} = 
p (1-\xi)^{-2}/ \left(1 - p + p \widehat{R}\right)$
coincides $\it exactly$ with the expression 
obtained earlier \cite{nous}
for the generating function
of the exponent of the survival probability of
an immobile target $A$ in presence of stochastically gated traps. 
The Tauberian theorem \cite{9} then insures that also in this general 
case of imperfect TR the inequality in Eq.(1) holds as $n \to \infty$.

To conclude, we have proven here that in the long-time limit
the survival probability of an $A$ particle performing random
walk on the sites of a $d$-dimensional lattice in presence of
randomly moving traps is less than or equal to the 
survival probability of an immobile $A$ particle in presence of randomly moving traps. 
This result holds for quite a general class of random walks as well as for perfect and imperfect
trapping reactions.

The authors thank J. Piasecki for 
fruitful discussions and
for pointing us on the Pascal's
assertion. We also acknowledge
discussions with A.J.Bray and R.A.Blythe.

\end{document}